\documentclass[a4paper]{revtex4}
\usepackage{graphicx}
\usepackage{fancyhdr}
\usepackage{amsmath}
\usepackage{color}
\newcommand{\Fb}{\bar{\Phi}}

\newcommand{\gbl}{g_{\rm B-L}}
\newcommand{\ubl}{U(1)$_{\rm B-L}$ }

\def\lrf#1#2{ \left(\frac{#1}{#2}\right)}
\def\lrfp#1#2#3{ \left(\frac{#1}{#2} \right)^{#3}}
\newcommand{\la}{\left\langle}  
\newcommand{\ra}{\right\rangle}
\def\beq{\begin{eqnarray}}
\def\eeq{\end{eqnarray}}
\def\s{\sigma}
\def\v{\varphi}
\def\GEV#1{10^{#1}{\rm\,GeV}}
\def\EQ#1{Eq.~(\ref{#1})}
\newcommand{\bear}{\begin{array}}  
\newcommand {\eear}{\end{array}}
\newcommand{\bea}{\begin{eqnarray}}   
\newcommand{\eea}{\end{eqnarray}}
\newcommand{\beal}{\begin{align}}   
\newcommand{\eeal}{\end{align}}
\newcommand{\lp}{\left(}
\newcommand{\rp}{\right)}

\begin{document}

\title{Inflation and Higgs}

%

\author{Fuminobu Takahashi}
\affiliation{Department of Physics, Tohoku University, Sendai 980-8578, Japan}

\begin{abstract}
We briefly review several Higgs inflation models and discuss their cosmological implications.
We first classify the inflation models according to the predicted value of the tensor-to-scalar 
ratio: (i) $r = {\cal O}(0.01-0.1)$, (ii) $r = {\cal O}(10^{-3})$, and (iii) $r \ll 10^{-3}$. 
For each case we study (i) the Higgs inflation with a 
running kinetic term, (ii) the Higgs inflation with a non-minimal coupling to gravity, and (iii) 
the $B-L$ Higgs inflation. In the last case we introduce supersymmetry to suppress the Coleman-Weinberg
corrections for successful inflation, and  derive the upper bound on the SUSY breaking scale. Interestingly,
the SUSY $B-L$ Higgs inflation requires the SUSY 
breaking scale of order ${\cal O}(100)$\,TeV to explain the observed spectral index.
 We briefly discuss a topological Higgs inflation which explains the origin of the standard model near-criticality.
We also mention the  possibility of Higgs domain walls and the gravitational waves emitted by the collapsing domain walls.
\end{abstract}

\maketitle

\thispagestyle{fancy}


\section{Introduction}
Inflation is strongly supported by various cosmological observations, especially the cosmic microwave background (CMB)
experiments~\cite{Ade:2015lrj}. The observed almost scale-invariant, adiabatic, and Gaussian density perturbation is consistent with 
the single-field slow-roll inflation paradigm. Yet, it remains unknown what the inflaton is. Here we focus on the possibility that
the standard model (SM) Higgs or other Higgs fields play the role of the inflaton. 

The Higgs inflation has an advantage that the successful reheating of the SM sector is automatic, and
the resultant reheating temperature tends to be rather high~\cite{GarciaBellido:2008ab}. On the other hand, if the inflaton is a gauge singlet,
one may have to introduce ad hoc couplings of the inflaton to the SM particles for successful reheating. This is not necessarily the case, however,
if the inflaton acquires a non-zero vacuum expectation value (VEV) after inflation.  Then, the inflaton generically decays into any sectors 
(including the SM and hidden sectors) through either direct Planck-suppressed couplings or
a mixing with the gravity sector~\cite{Endo:2006qk,Endo:2007ih,Endo:2007sz,Watanabe:2006ku}\footnote{In supergravity, the decay into the 
SUSY breaking sector leads to non-thermal gravitino production~\cite{Endo:2007ih,Endo:2007sz}.}. In this case, the reheating temperature
tends to be lower, as the decay rate is Planck-suppressed. 
 
 In the following we consider three  Higgs inflation models which predict the tensor-to-scalar ratio,  $r = {\cal O}(0.01-0.1)$,
$r = {\cal O}(10^{-3})$, and $r \ll 10^{-3}$. For each case we study (i) the Higgs inflation with a 
running kinetic term, (ii) the Higgs inflation with a non-minimal coupling to gravity, and (iii) 
the $B-L$ Higgs inflation.

\section{Various Higgs Inflation Models}

\subsection{Higgs Inflation with a running kinetic term}
Now let us consider an inflation model where the SM Higgs field plays the role of the inflaton. 
For successful inflation, the Higgs potential must become flatter at large field values. 
Here we consider a possibility that the form of the kinetic term changes so as to make
the potential sufficiently flat. This is known as the running kinetic inflation~\cite{Takahashi:2010ky,Nakayama:2010kt}.

The basic idea of the running kinetic inflation is very simple. 
 Let us consider a scalar field $\phi$ with the following Lagrangian,
\begin{eqnarray}
\label{example}
{\cal L} &=& \frac{1}{2} \left(1+ \xi \phi^2 \right) (\partial \phi)^2 - V(\phi),
\end{eqnarray}
where $\xi$ is a positive coefficient larger than unity, and $V(\phi)$ is the inflaton 
potential. Here and in what follows we adopt the Planck units
in which the reduced Planck scale $M_P\simeq 2.4\times 10^{18}\,$GeV is set to be unity.
Due to the dependence of the kinetic term on $\phi^2$, the canonically normalized field
at  $\phi \gtrsim 1/\sqrt{\xi}$ is given by ${\hat \phi} \sim \sqrt{\xi} \phi^2$.
As the kinetic term grows, the potential becomes
flatter  in terms of the canonically normalized field. For instance, the quartic potential, $V(\phi) \sim \phi^4$, becomes the quadratic one, 
$V({\hat \phi}) \sim {\hat \phi}^2/\xi$,  at large field values. For a different form of the kinetic term, a fractional-power potential can also be
realized.  This is the essence of the running kinetic inflation.
The running kinetic inflation can be easily implemented in
 supergravity and the cosmological implications were studied in Refs.~\cite{Takahashi:2010ky,Nakayama:2010kt}.

The above argument can be straightforwardly applied to the SM Higgs field, and the SM Higgs can
drive quadratic or fractional-power chaotic inflation~\cite{Nakayama:2010sk}. In order to build sensible inflation models, we need to have a good control of 
the scalar potential over large field values. Also, it  is desirable to understand the large value of  $\xi \gg 1$ 
in terms of symmetry. 
To this end,  we introduce an approximate shift symmetry on the absolute square of the SM 
Higgs field $H$~\cite{Takahashi:2010ky,Nakayama:2010kt,Nakayama:2010sk}:
\begin{equation}
	|H|^2 \to |H|^2 + C,
	\label{ss}
\end{equation}
where $C$ is a real transformation parameter and the SU(2)$_L$ indices are omitted.
The shift symmetry becomes apparent at high energy scales, while it is explicitly broken and therefore 
somewhat hidden at low energy scales.
The Lagrangian  at high energy scales is given by 
\begin{equation}
	\mathcal L =  \frac{1}{2} \left(\partial_\mu |H|^2\right)^2+ \epsilon |D_\mu H|^2 
	 - \lambda \left(|H|^2 - \frac{v^2}{2}\right )^2 + \cdots,
	\label{L}
\end{equation}
where $\epsilon$ and $\lambda$ are coupling constants, and $D_\mu$ denotes the covariant derivative. 
The first term in (\ref{L}) respects the shift symmetry, which is explicitly broken by the second and third terms, and so,
we  expect $\epsilon, \lambda \ll 1$. Note that, while the second term provides the usual kinetic term for the SM Higgs
in the low energy,  the first term provides the kinetic term for $|H|^2$ at large field values.
In the unitary gauge, the relevant interactions are
\begin{equation}
	\mathcal L = \frac{1}{2}\left(\epsilon + h^2 \right) (\partial h)^2 - \frac{\lambda}{4}\left(h^2- v^2\right)^2,
\end{equation}
where $h$ denotes the physical Higgs boson, and we have omitted the gauge and Yukawa interactions as they are irrelevant
during inflation. 
Thus, the largeness of $\xi$ in (\ref{example}) is due to the smallness of $\epsilon$, i.e., the fact that the usual kinetic term
breaks the shift symmetry (\ref{ss}).

For large field value $h \gg \sqrt{\epsilon}$, we can rewrite the Lagrangian in terms of
the canonically normalized field $\hat h \equiv h^2/2$ as
\beq
	\mathcal L \simeq \frac{1}{2}(\partial \hat h)^2 - \lambda \hat h^2.
\eeq
The Planck normalization on the density perturbation fixes $\lambda = m^2/2 \simeq 2 \times 10^{-11}$~\cite{Ade:2015lrj}.
Thus, the chaotic inflation with quadratic potential can be realized by the SM Higgs field with the running kinetic term.
The predicted values of $n_s$ and $r$ are
\begin{align}
n_s -1& \simeq -0.033 \lrf{60}{N}, \\
r&\simeq 0.13 \lrf{60}{N},
\end{align}
where $N$ is the e-folding number.
Note that all the interactions of the (canonically normalized) Higgs field are suppressed and the system approaches the
free field theory as $h$ increases. 

For small field value $h \ll \sqrt{\epsilon}$,
the Lagrangian is reduced to the usual one for the SM Higgs field,
\beq
	\mathcal L \simeq \frac{1}{2}(\partial \tilde h)^2 - \frac{\tilde\lambda}{4}\left(\tilde h^2 - \tilde v^2\right)^2,
\eeq
where  we have defined  $\tilde h \equiv \sqrt{\epsilon} h$,
$\tilde\lambda \equiv \lambda / \epsilon^2$ and $\tilde v \equiv \sqrt{\epsilon}v$. 
In order to explain  the  correct electroweak scale and the 125\,GeV Higgs boson mass,
we must have $\tilde v =246\,$GeV and  $\tilde\lambda \simeq 0.13$.
The SM Yukawa interactions are also obtained in the low energy if we add the Yukawa interactions
with suppressed couplings in \EQ{L}~\cite{Nakayama:2010sk}. That is to say, the low-energy
effective theory coincides with the SM at ${\tilde h} \lesssim \epsilon$, while the theory asymptotes a
 free theory for a massive scalar ${\hat h}$ at ${\tilde h} \gtrsim \epsilon$. The transition from one phase to the other takes
place at an intermediate scale $\sim \GEV{13}$ or above,  whose precise value depends on the running of the
quartic coupling ${\tilde \lambda}$. As the transition takes place at an intermediate scale, 
the constraint on the top quark mass is much milder compared to the Higgs inflation with a non-minimal coupling to gravity.

\subsection{Higgs Inflation with a non-minimal coupling to gravity}
The Higgs inflation with a non-minimal coupling to gravity~\cite{Bezrukov:2007ep} is one realization of the so called 
induced gravity inflation model~\cite{Salopek:1988qh}. The inflaton dynamics as well as it implications for the Higgs phenomenology
has been studied extensively in the literature, and we are not going to
repeat the analysis here. The point is that, once one introduces a non-minimal coupling to gravity, 
\begin{align}
S &=\int d^4 x \sqrt{-g} \left[  \frac{M_P^2+\xi h^2 }{2}   R
+ \frac{1}{2} \partial_\mu h \partial^\mu h 
- \frac{\lambda}{4}\left( h^2 - v^2 \right)^2 \right],
\end{align}
the Higgs potential becomes flat at sufficiently
large field values in the Einstein frame as
\begin{align}
V(\phi) &
\simeq \frac{\lambda M_P^4}{4 \xi^2}\lp 1+ e^{-\sqrt{\frac{2}{3}} \frac{\phi}{M_P}}\rp^{-2}.
\end{align}
Here $\phi$ is the canonically normalized inflaton field at large field values, and it is related to $h$ as
\beal
\phi & \simeq \sqrt{6} M_P \ln \lp \sqrt{\xi}h \over M_P \rp
\end{align}
for  $h \gg 1/\sqrt{\xi}$. The predicted values of $n_s$ and $r$ are
\begin{align}
n_s -1& \simeq -0.033 \lrf{60}{N}, \\
r&\simeq 3.3 \times 10^{-3} \lrfp{60}{N}{2}.
\end{align}

Lastly we mention the relation with the running kinetic inflation. As pointed out in Ref.~\cite{Bezrukov:2014bra},
the running kinetic inflation studied in the previous subsection can be realized
if one chooses a specific form of the non-minimal coupling:
\begin{align}
M_P^2 R \sqrt{1+\xi |H|^2/M_P^2},
\end{align}
which  leads to the quadratic chaotic inflation.
Therefore, the two inflation models are related to each other to some extent, even though their construction is 
quite different.

\subsection{B-L Higgs Inflation}
Now let us consider a model in which a Higgs boson $\varphi$ responsible for the breaking of
U(1)$_{\rm B-L}$ symmetry plays the role of the inflaton.   If the inflaton potential is sufficiently flat around the origin, 
and if the inflaton initially sits in the vicinity of the origin, the inflation takes place. As $\varphi$ is charged under the \ubl
symmetry, the inflaton potential receives a radiative correction from
the gauge boson loop.  The general form of the CW effective potential
is given by~\cite{Coleman:1973jx}
\begin{equation}
V_{\rm CW}\; =\; \frac{1}{64\pi^2}
\sum_i (2 S + 1) (-1)^{2S} M_i^4(\varphi) \ln\lrf{M_i^2(\varphi)}{\mu^2},
  \label{CW}
\end{equation}
where $\mu$ is the renormalization scale, and 
 the mass eigenvalues of the particles coupled to $\varphi$ are represented by $M_i (\varphi)$.
Since the mass of the U(1)$_{\rm B-L}$ gauge boson is given by $m_{\rm
  GB} = \sqrt{2} \gbl q_\varphi \la \varphi \ra$, the inflaton
potential receives the CW correction as
\beq
V_{\rm CW, gauge}(\s) \;=\; 
\frac{3}{64\pi^2} \gbl^4 q_\varphi^4  \s^4\, \ln\left(\frac{\gbl^2 q_\v^2  \s^2}{\mu^2} \right),
  \label{CW1}
\eeq
where $\gbl$ represents the gauge coupling of U(1)$_{\rm B-L}$,
$q_\varphi$ is the \ubl charge of $\varphi$, and $\sigma$ denotes the
radial component of $\varphi$, $\sigma \equiv \sqrt{2} |\varphi|$.

It is well known that the CW potential arising from the gauge boson
loop makes the effective potential so steep that the resultant density
perturbation becomes much larger than the observed
one~\cite{Starobinsky:1982ee,Hawking:1982cz,Guth:1982ec}.  One plausible way to solve the problem
is to introduce SUSY~\cite{Ellis:1982ed}.  In the exact SUSY limit,
contributions from boson loops and fermion loops are exactly canceled
out.  However, if SUSY is broken, we are left with non-vanishing CW
corrections, which are estimated below, based on Refs.~\cite{Nakayama:2011ri,Nakayama:2012dw}.

In SUSY, two U(1)$_{\rm B-L}$ Higgs bosons are required for anomaly
cancellation.  Let us denote the corresponding superfields as
$\Phi(+2)$ and ${\bar \Phi}(-2)$ where the number in the parenthesis
denotes their B$-$L charge.  The $D$-term potential vanishes along the
$D$-flat direction $\Phi {\bar \Phi}$, which is to be identified with
the inflaton. Actually, a linear combination of the lowest components
of $\Phi$ and $\bar{\Phi}$ corresponds to $\v$. We can simply relate
$\Phi$ and $\Fb$ to $\v$ as $|\Phi| = |\Fb| = |\v|/\sqrt{2}$. The \ubl
charge of $\v$ is set to be $q_\v = 2$ in the following.

The gauge boson has mass of $m_{S}^2 = \gbl^2 q_\v^2 \s^2$, where $\sigma
\equiv \sqrt{2} |\v|$.  On the other hand, there are additional
fermionic degrees of freedom, the U(1)$_{\rm B-L}$ gaugino and
higgsino, whose mass eigenvalues are given by $m_F = \gbl q_\v \s \pm
\frac{1}{2} M_\lambda$, where $M_\lambda$ denotes the soft SUSY
breaking mass for the \ubl gaugino.  Because of the SUSY breaking mass
$M_\lambda$, the CW potential does not vanish and the inflaton
receives a non-zero correction to its potential.  Inserting the field
dependent masses into the CW potential (\ref{CW}), and expanding it by
$M_\lambda/ (g q_\v \s)$, we find
\begin{equation}
	V_{\rm CW, gauge}^{\rm susy}(\s) \simeq - \frac{3 \gbl^2}{8\pi^2}
	\lrfp{q_\v}{2}{2} M_\lambda^2 \s^2
	\ln \left(\frac{\gbl^2 q_\v^2 \s^2}{\mu^2} \right),
	\label{CW2}
\end{equation}
where we have also taken into account of the inflaton as well as the scalar perpendicular to
the D-flat direction. 
Thus, in the presence of SUSY, the CW potential becomes partially
canceled and the dependence of the inflaton field has changed from
quartic to quadratic as long as $M_\lambda \ll \gbl q_\v \sigma$, in
contrast to the result of Ref.~\cite{Ellis:1982ed}.

For successful inflation, we require the curvature of the CW potential
(\ref{CW2}) to be at least one order of magnitude smaller than $H_{\rm
  inf}$ for $\s \lesssim \s_{\rm end}$.  Here
  $H_{\rm  inf}$ is the Hubble parameter during inflation, and 
   $\s_{\rm end}$ is the point where the slow-roll
condition breaks down and the inflation ends.  Therefore, we obtain
the following constraint on the soft SUSY breaking mass for the \ubl
gaugino:
\begin{equation}
	\gbl M_\lambda \; \lesssim \;  H_{\rm inf}.
	\label{bound}
\end{equation}
For the gauge coupling of order unity, this bound reads 
$M_\lambda \; \lesssim \;  H_{\rm inf}$.

The \ubl Higgs boson is also coupled to the right-handed neutrinos to
give a large Majorana mass. Barring cancellations, similar argument leads to 
the upper bound on the soft SUSY breaking mass of the right-handed sneutrino,
\beq
\sum_i y_{\v,i}^2 m_{\tilde N,i}^2\;\lesssim\;  H_{\rm inf}^2,
	\label{bound2}
\eeq
where $y_{\v,i}$ denotes the coupling of of the $B-L$ Higgs to the $i$-th
right-handed neutrino.

In the gravity mediation, $M_\lambda$, $m_{\tilde N,i}$ as well as the
soft SUSY masses for the SUSY SM particles are considered to be comparable
to the gravitino mass $m_{3/2}$.  On the other hand, in anomaly
mediation~\cite{Giudice:1998xp}, they may be suppressed compared to
the gravitino mass, but for a generic form of the K\"ahler potential,
$m_{\tilde N}$ and the sfermion masses are comparable to the gravitino
mass. 

To summarize, successful $B-L$ Higgs inflation places a robust upper bound on the soft SUSY breaking parameter
of the \ubl gaugino and the right-handed sneutrinos.  In particular, for
the gauge and Yukawa couplings of order unity, both $M_\lambda$
and $m_{\tilde N,i}$ should be smaller than $H_{\rm
  inf}$. If $M_\lambda$ and/or $m_{\tilde N,i}$ are
comparable to the gravitino mass, we obtain
\beq
       	     m_{3/2} \;\lesssim\;  H_{\rm inf},        
	      \label{mH}  
\eeq
which relates the inflation scale to the SUSY breaking. 

The tree-level inflaton potential can be written as
\beq
V(\sigma) \;\simeq\; v^4  - \frac{1}{2} k v^4  \sigma^2
- \frac{g }{2^{2n-1}} v^2 \sigma^{2n} + \frac{g^2}{2^{4n}} \sigma^{4 n},
\label{Vinf}
\eeq
where $k \ll 1$ is required to avoid the so called $\eta$-problem.
The inflation scale varies from $10^6$\,GeV ($n=2$)
to $\GEV{10}$ or heavier ($n\geq3$).  Interestingly, in the case of $n=2$,
 the VEV of the inflaton is very close to the see-saw scale
$\sim \GEV{15}$ suggested by the neutrino oscillation data.  Therefore, the successful SUSY $B-L$ Higgs
inflation together with the seesaw mechanism require the SUSY breaking scale smaller than  or comparable
to ${\cal O}(10^5)$\,GeV. The non-thermal leptogenesis also works.
As shown in Ref.~\cite{Nakayama:2012dw}, the upper bound must be
saturated to realize the observed spectral index $n_s \sim 0.96$. Therefore,   ${\cal O}(100)$\,TeV SUSY  and
the $125$\,GeV Higgs boson mass may be the outcome of the SUSY $B-L$ Higgs inflation.

\section{Topological Higgs Inflation and Higgs domain walls}
The measured Higgs boson mass about $125$\,GeV implies that the SM could be valid all the way up to the Planck scale. 
If so, the Higgs potential  {may have} another minimum around the Planck scale,  {depending on the top quark mass}~\cite{Branchina:2014rva}.
If the extra minimum is the global one, our electroweak
vacuum is unstable and  decays through quantum tunneling processes with a finite lifetime.
 {In contrast, nature may realize} a critical situation where the two minima are degenerate in energy.
Froggatt and Nielsen focused on this special case, the so-called  Higgs criticality; they  provided a theoretical argument to support this case,  {the multiple point criticality principle~\cite{Froggatt:1995rt,Nielsen:2012pu}}.

It was pointed out in Ref.~\cite{Hamada:2014raa} that the near-criticality can be understood if the Universe experiences
eternal topological inflation~\cite{Linde:1994hy, Linde:1994wt,Vilenkin:1994pv} induced by the Higgs field
at a very early stage. The condition for the topological inflation is only marginally 
satisfied in the SM, but it can be readily satisfied if one extends the SM
by introducing heavy right-handed neutrinos and/or a non-minimal coupling to gravity. 

The topological Higgs inflation may be thought of as one of the variants of the Higgs inflation,
but it is different in the following aspects. First, the topological inflation is free of the initial
condition problem. If the Universe begins in a chaotic state at an energy close to the Planck scale,
the Higgs field may take various field values randomly up to the Planck scale or higher~\cite{Linde:1983gd}. 
As the Universe expands, the energy density decreases and the Higgs field finds itself
either larger or smaller than the critical field value corresponding to the local maximum, and gets
trapped in one of the two degenerate vacua with a more or less equal probability. 
This leads to  formation of domain walls separating the two vacua. Interestingly,
then, eternal inflation could take place inside the domain walls, if the thickness of the domain walls is
greater than the Hubble radius~\cite{Linde:1994hy,Vilenkin:1994pv}. 
In this sense, no special fine-tuning of the
initial position of the inflaton is necessary for the inflation to take place. Specifically, 
the topological inflation  occurs if the two minima are separated by more than the Planck 
scale,  which was also confirmed by numerical calculations~\cite{Sakai:1995nh}. 
Secondly, the magnitude of density perturbations generated
by topological Higgs inflation tends to be too large to explain the observed CMB temperature fluctuations. 
This is similar to the problem of the GUT Higgs inflation~\cite{Starobinsky:1982ee,Hawking:1982cz,Guth:1982ec}.
We need therefore another inflation after the end of the topological Higgs inflation. Thus, the role of the topological Higgs inflation is to continuously 
create sufficiently flat Universe, solving the so-called longevity problem of inflation with a Hubble parameter much 
smaller than the Planck scale~\cite{Linde:1983gd,Izawa:1997df}; the Universe must be sufficiently flat and
therefore long-lived so that the subsequent slow-roll inflation with a much smaller Hubble parameter can take place.

Lastly let us mention a possibility of the Higgs domain walls which are formed after inflation and collapse, emitting sizable 
gravitational waves~\cite{Kitajima:2015nla}.
Let us consider the following Higgs potential lifted by new physics at some high energy scale,
\beq
	V_H = \frac{1}{4} \lambda(\varphi) \varphi^4 + \frac{\varphi^6}{\Lambda^2}.
	\label{eq:higgs_pot}
\eeq
where $\varphi$ is the Standard Model Higgs scalar field, $\lambda(\varphi)$ is a scale-dependent self coupling constant and $\Lambda$ is a cutoff scale
for the dimension six operator.
We have neglected the quadratic term of order  the electroweak scale as we are interested in the behavior of the Higgs potential at high energy.

\begin{figure}
 \includegraphics[width=.40\textwidth]{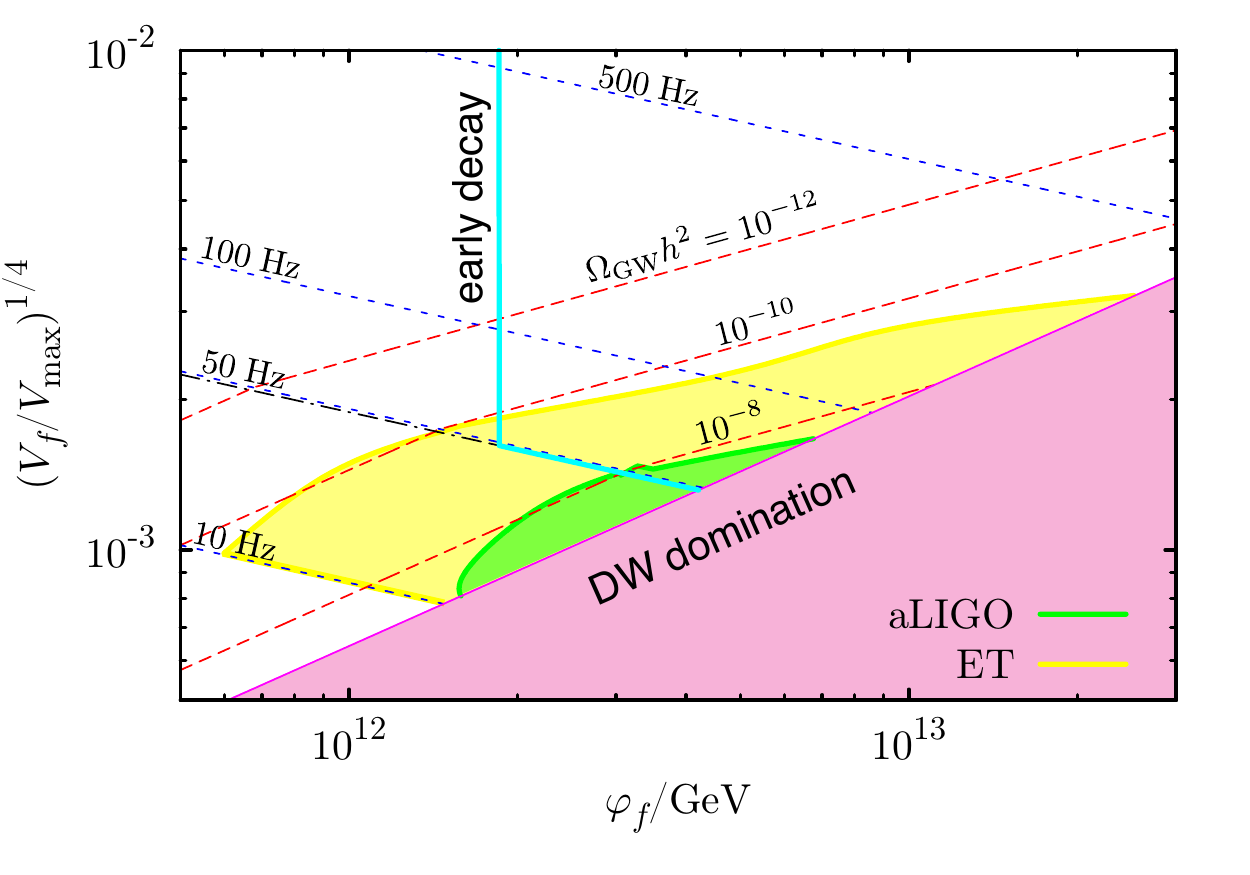}
 \caption{\label{asdf}
 The gravitational abundance and the sensitivity reach of the gravitational wave experiments. The horizontal axis
 is the position of the false vacuum, and the vertical axis is the ratio of the energy density of the false vacuum 
 to that of the local maximum. The figure is from Ref.~\cite{Kitajima:2015nla}.}
 \end{figure}

The effective potential can be lifted by higher dimensional operators such as
the last term in the right-hand-side in Eq.~(\ref{eq:higgs_pot}). In this case there are two potential minima; one is at the EW scale $\varphi = v_{\rm EW} $, and the other 
at a much higher scale $\varphi = \varphi_f$. Depending on the size of the higher dimensional operator, the high-scale minimum can be a local or
global minimum. In particular, our main interest lies in the case when the two minima are quasi-degenerate and the EW vacuum is slightly
energetically preferred:
\bea
&V_H(v_{\rm EW}) \approx V_H(\varphi_f) \approx 0, \\
&V_H(v_{\rm EW}) <V_H(\varphi_f).
\eea
If the Higgs field acquires a sufficiently large quantum fluctuations during inflation,
 both vacua may be populated in different patches of the Universe, leading to domain wall formation 
after inflation. In a later Universe,  the EW vacuum will be selected after domain wall annihilation. 
The domain walls subsequently annihilate, emitting a sizable amount of gravitational waves.
See Fig.~\ref{asdf} where the abundance of the gravitational waves as well as the sensitivity regions of
the advanced LIGO and ET are shown.

\begin{acknowledgments}
This work was supported by  JSPS Grant-in-Aid for
Young Scientists (B) (No.24740135),  Scientific Research (A) (No.26247042), 
Scientific Research (B) (No.26287039), and the Grant-in-Aid for Scientific Research on Innovative Areas (No.23104008 [FT]). 
This work was also supported by World Premier International Center Initiative (WPI Program), MEXT, Japan.
\end{acknowledgments}

\bigskip 

\end{document}